\documentclass{article}

\usepackage[utf8]{inputenc} 
\usepackage[T1]{fontenc} 
\usepackage{cite} 

\usepackage{extpfeil}
\usepackage{fancyvrb}
\usepackage{listings}
\usepackage{tipa}

\usepackage{hyperref}
\usepackage{graphicx}

\newcommand{\cabernet}{\textsc{CABeRNET}}

\begin{document}
\title{{CABeRNET}: a Cytoscape app for Augmented Boolean models of gene
  Regulatory NETworks}

\author{Andrea Paroni, Alex Graudenzi, Giulio Caravagna,\\
Chiara Damiani, Giancarlo Mauri, Marco Antoniotti\\[2mm]
Dipartimento di Informatica, Sistemistica e Comunicazione\\
Universit\`{a} degli Studi di Milano Bicocca\\
Milan, \textsc{Italy}}

\maketitle

\begin{abstract} 
\noindent
\textbf{Background.}
Dynamical models of gene regulatory networks (GRNs) are highly
effective in describing complex biological phenomena and processes,
such as cell differentiation and cancer development. Yet, the
topological and functional characterization of real GRNs is often
still partial and an exhaustive picture of their functioning is
missing.

\noindent
\textbf{Motivation.}
We here introduce \cabernet{}, a Cytoscape app for the generation,
simulation and analysis of Boolean models of GRNs, specifically
focused on their augmentation when  a only partial topological and
functional characterization of the network is available. By generating
large ensembles of networks in which user-defined entities and
relations are added to the original core,  \cabernet{} allows to
formulate hypotheses on the missing portions of real networks, as well
to investigate their generic properties, in the spirit of complexity
science.

\noindent
\textbf{Results.} 
\cabernet{} offers a series of innovative simulation and modeling
functions and tools, including (but not being limited to) the
dynamical characterization of the gene activation patterns ruling cell
types and differentiation fates, and sophisticated robustness
assessments, as in the case of gene knockouts. The integration within
the widely used Cytoscape framework for the visualization and analysis
of biological networks, makes \cabernet{} a new essential instrument
for both the bioinformatician and the computational biologist, as well
as a computational support for the experimentalist.  An example
application concerning the analysis of an augmented T-helper cell GRN
is provided.
\end{abstract}

\section*{Introduction}

Consistently with the  increasing availability of  \emph{big data}
regarding biological systems, is the need of \emph{mathematical} and
\emph{computational models} aimed at their effective analysis and
interpretation \cite{Kitano2001,Kitano2002}. To this end, in the
spirit of \emph{statistical physics} and \emph{complex systems
  science}, even  simplified and  abstract models have proven
effective in representing complex biological phenomena, with specific
focus on the \emph{emergent} dynamical behaviours and the so-called
\emph{generic} or \emph{universal} properties
\cite{Kauffman_1993,Kaneko2006}.

In the context of genomics data, one of the best examples is that
provided by Boolean models of \emph{gene regulatory networks} (GRNs),
which have repeatedly proved fruitful in describing key properties of
real systems, as well as in providing cues and hints for  wet-lab
experiments (see,
e.g.,\cite{Kauffma1995,shmulevich2002probabilistic,shmulevich2002boolean,Kauffman2003,Serra2004,Kauffman2004,Shmulevich2005,Ramo2006,Serra2007,serra2008simulation,cheng2010analysis}). The
simulation of partially characterized regulatory architectures with a
Boolean approach, in particular, has recently gained attention (see,
e.g.,
\cite{Sanchez1997,akutsu1999identification,Sanche2001,Albert2003,li2004yeast,chaos2006genes,faure2006dynamical,klamt2006methodology,davidich2008boolean}). A
first motivation lies in the inherently ``dynamical'' nature of  gene
(de)regulation processes, and in the clear limitations of a ``static''
analysis  capturing only a partial picture of such complex
processes. For example, a structural analysis of the genomic
interactions might preclude to determine the influence of a
\emph{target-selective therapy} on the overall GRN interplay ruling
\emph{tumorigenesis}.

Moreover, as compared to continuous models, such as ODE-based or
stochastic models (see \cite{karlebach2008modelling} for a review on
GRN modeling), the Boolean abstraction allows for a clear and
effective characterization of the  \emph {gene activation patterns}
(or \emph{attractors} in the terminology of dynamical systems)
characterizing the different phenotypic functions, such as cell types,
modes and fates, under the metaphor of ``emergent collective
behaviours''\cite{Kauffman_1993,Huang2000,huang2005cell,Forgacs2006,Ribeiro2007,Huang_2009,Canham2010,felli2010hematopoietic,furusawa2012dynamical,cheng2013biomolecular,nikolov2014tumors}.
In this context, phenomena such as  tumorigenesis can be explained as
rare emergent pattern, triggered by  \emph{signals}, stochastic
fluctuations and \emph{biological noise} (see,
e.g.,\cite{Eldar2010,tsimring2014noise}).

In line with this approach, we here continue on a recent research
strand that has involved the development  of simulation and analysis
tools for dynamical Boolean GRNs (see, e.g., \cite{Antoniotti2013}).
The foundations of our theoretical framework lay in the seminal work
by Stuart Kauffman on \emph{Random Boolean Networks} (RBNs)
\cite{Kauffman69_a,Kauffman69_b} and, more recently, on the dynamical
model of cell differentiation
introduced in~\cite{serra2010,Villani2011} and based on \emph{Noisy
  Random Boolean Networks} (NRBNs).  In this theoretical framework,
cell types are associated to dynamical gene activation patterns and
differentiation to cell-specific noise-resistance mechanisms, the
underlying hypothesis being that such mechanisms refine along with
differentiation stages (see \cite{Hoffman2008} and references
therein). In the Background Section the main features of this approach
are outlined.

In this paper we introduce \cabernet{}, a Cytoscape \cite{Shannon2003}
app for the generation, import, simulation and analysis of Boolean
models of GRNs. With respect to similar approaches, e.g.,
\cite{le2011netds} or  the earlier work by our group
\cite{Antoniotti2013}, \cabernet{} also allows to \emph{augment}
partially-characterized GRNs by adding user-defined entities and
relations.  The underlying motivation is that, despite the increasing
knowledge on gene regulation in real organisms, the topological and
functional characterization of real networks is stil far from being
comprehensive, thus preventing to capture the complexity of the
overall interplay: \cabernet{} is conceived to randomly generate the
missing portions of the partially-characterized GRNs, in order to
obtain \emph{ensembles} of augmented GRNs that share the common core
and other structural and functional parameters, hence allowing to test
hypotheses on the yet unknown missing portions of real
networks. Besides, in line with the complex systems approach, the
generic properties of such ensembles can be inferred and investigated,
possibly providing an ensemble-level understanding of yet unraveled
GRN phenomena. In addition, \cabernet{} guides the user through
various robustness analyses of the GRNs, which can be matched against
genomic experimental data. Notice that \cabernet{} can also generate
completely random GRNs with defined structural   and functional
constraints, as well as simulate the dynamics of completely
characterized GRNs. In the Implementation Section the main features of
\cabernet{} are described.

As a proof of principle, in the Results Section we present the
augmentation of the T-helper cell signaling network and describe the
analysis of its dynamics, with particular focus on the emergent
differentiation scheme and its robustness against the knockout of
specific genes.  Conclusions are presented in the last Section.

\section*{Background: the GRN model}

In \cabernet{}, GRNs are represented as Noisy Random Boolean Networks
\cite{Villani2011}.  A classical RBN is a graph composed of $n$
Boolean nodes associated with binary variables $\sigma_{i}\in\{0,1\}$,
representing the activation of a gene: if $\sigma_{i}=1$ the $i-$th
gene synthesizes its product (i.e., proteins or RNAs), otherwise it
does not. Each node is connected to $k_{i} \leq n-1$  nodes which
represent its regulatory inputs, i.e., those genes  influencing the
activation of the $i-$th gene. The way in which regulation takes place
is via a node-specific Boolean function $f_{i}$ of the regulatory
inputs: the value of $\sigma_{i}$ at time $t+1$ is determined as
$\sigma_{i}(t+1)=
f_{i}(\sigma_{j_{1}}(t),\sigma_{j_{2}}(t),\ldots,\sigma_{j_{k_{i}}}(t))$.
The dynamics of canonical RBNs is synchronous,  deterministic and
follows discrete time steps.  Thus, it eventually ends up in (at least
unitary) state cycles, termed \emph{attractors}, which represent the
emerging gene activation patterns displayed by the GRN.  Remarkably,
by varying the structural features of the networks, different
dynamical regimes appear, ranging from ordered to disordered behaviors
(see \cite{Kauffman_1993}).

\begin{figure}[ht]
  \begin{center}
    \includegraphics[width=0.9\textwidth]{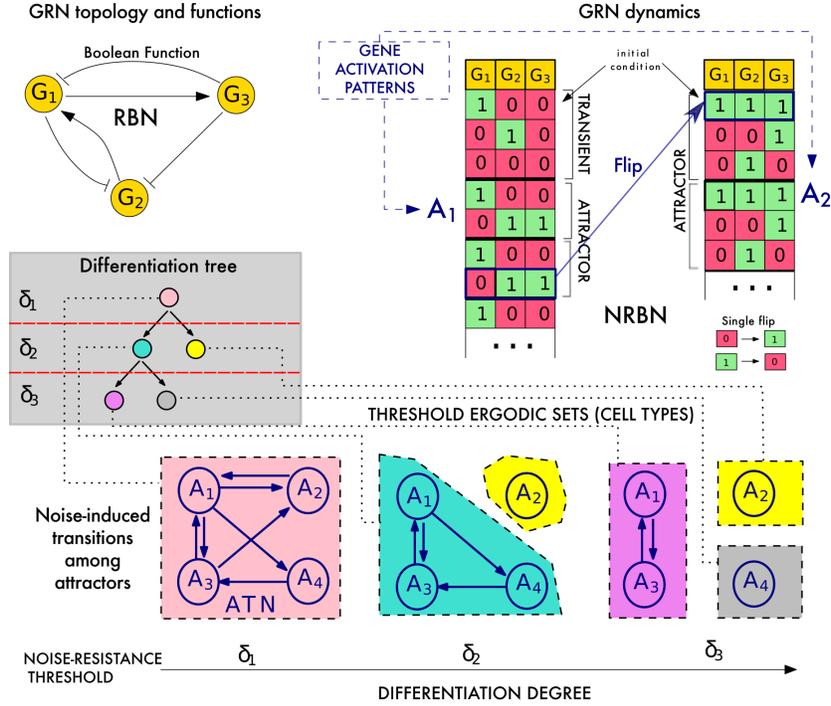}
  \end{center}
  \caption{\footnotesize{\bf Simplified representation of  the GRN
      model in \cabernet{}}. (a) Example RBN with $3$ genes (nodes)
    and  edges representing regulatory interactions (via node-specific
    Boolean functions, not shown). (b) Example  dynamics to highlight
    network's attractors, modeling gene activation patterns $A_1$,
    $\ldots$, $A_5$, and possible transitions among them induced by
    noise  (i.e., single flips). (c) The transitions yield an
    Attractor Transition Network that generates $5$ cellular types
    when $3$  thresholds, $\delta_i$, $i=1,2,3$ are evaluated to asses
    the  corresponding Threshold Ergodic Sets. In this approach, where
    the efficiency of noise-control mechanisms  is related to
    differentiation types, stem cells (pink), intermediate stages
    (light blue) and fully differentiated cells (yellow, purple and
    grey) emerge. The corresponding  differentiation tree, is shown
    (Figure modified from \cite{Graudenzi2014}).}
\label{fig:NRBN}
\end{figure}

In \cite{Villani2011} Villani et al. introduced a form of
stochasticity in RBNs, hence speaking of ``Noisy RBNs''. NRBNs allow
for the transitions among attractors as a consequence of random
``flips'' of $\sigma_i$'s value, i.e., random modifications of the
node's activation value lasting a defined time span. Thus, with this
approach it is possible to determine the so-called \emph{Attractors
  Transition Network} (ATN, or \emph{matrix}, ATM), i.e., a
\emph{stability matrix}  displaying the probability of a gene
activation pattern to emerge in response to a perturbation in another
pattern (noise-induced transition). A \emph{noise threshold} is then
introduced to exclude transitions unlikely to occur for a cell in a
certain differentiation stage. Therefore, multiple thresholds allow to
identify distinct \emph{threshold-dependent} ATNs, each one
representing the stability of the GRN's activation patterns
characterizing a specific differentiation stage (e.g.,
toti-/multi-potents stem cells, progenitors, differentiated cells,
etc). The exact mathematical definition of such patterns is given as
\emph{Threshold Ergodic Sets} (TESs), which: $i)$ are \emph{strongly
  connected components} in the threshold-dependent ATN, $ii)$  and
have no outward transitions outside the components' set. These
structures are hierarchical, thus can be naturally mapped to
\emph{cell types} yielding to emergent \emph{differentiation
  trees}. Such trees mimic the metaphor that less differentiated
stages have less refined noise-control mechanisms, resulting hence in
TES associated to lower thresholds (see, e.g.,
\cite{Hu1997,Hayashi2008,Furusawa2009}). In general, the number of
patterns observable in a certain cell type decreases progressively
along with the differentiation stage (see Fig.~\ref{fig:NRBN} for a
simplified representation of the model).

Notice that the approach is general, it does not refer to any specific
organism, and has been shown to be suitable to describe $i)$
different degrees of differentiation, i.e., from toti-/multi-potent
stem cells to intermediate states, to fully differentiated cells;
$ii)$ the stochastic differentiation process, in which a population of
toti-/multi-potent cells stochastically generates progenies of
distinct types; and $iii)$ the induced pluripotency phenomenon,
according to which fully differentiated cells can revert to a
pluripotent stage through the perturbation of some key genes
\cite{Yamanaka2009}.

\section*{Implementation}

\begin{figure}[ht!]
  \begin{center}
    \includegraphics[width=0.95\textwidth]{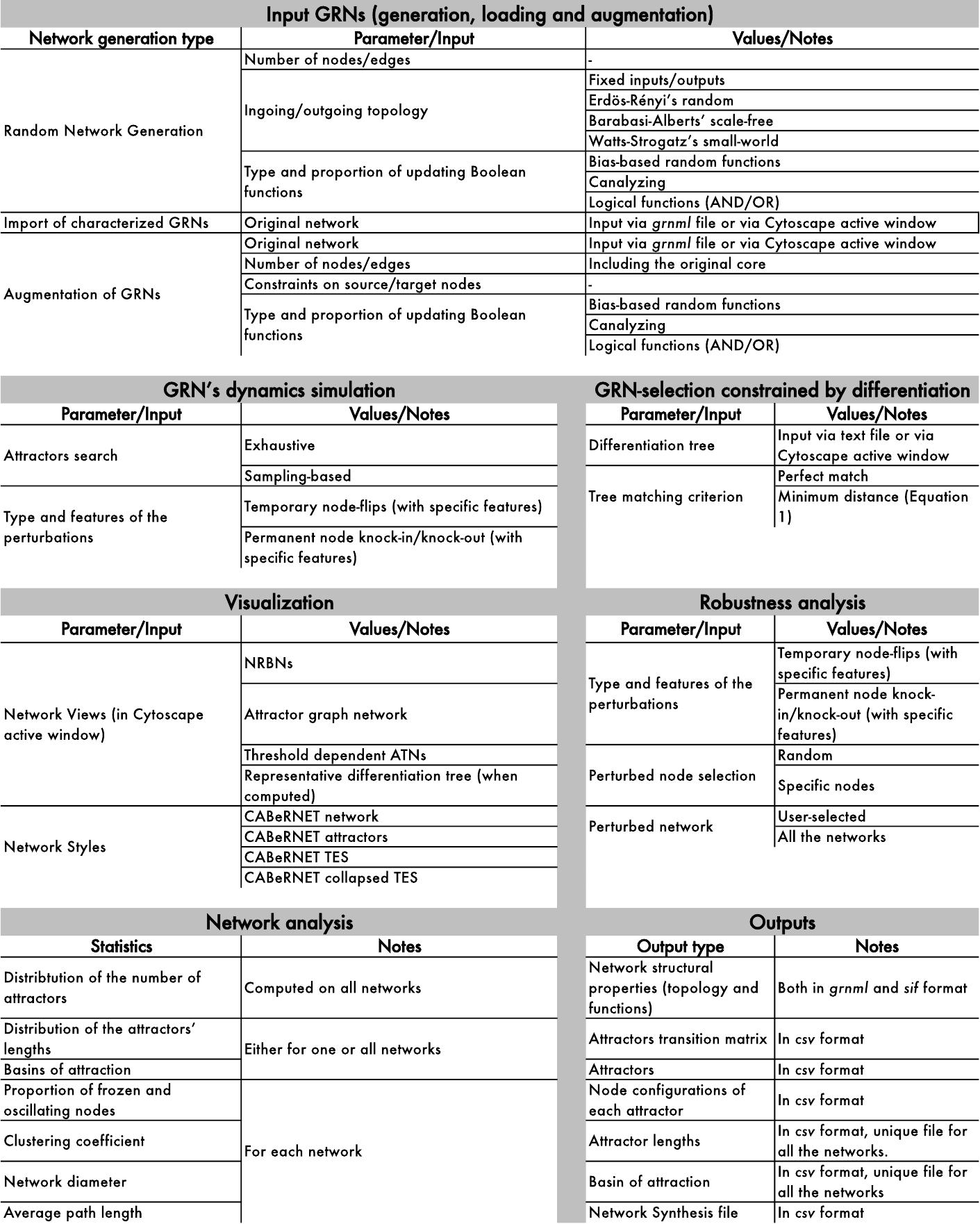}
  \end{center}
  \caption{\footnotesize{\bf Main functionalities and parameters of
      \cabernet{}}. A schematic representation of the various
    functions and parameters  of \cabernet{} is provided, as
    explicitly described in the Implementation Section in the main
    text. For a thorough explanation please refer to the user manual
    (see Availability). }
  \label{fig:table}
\end{figure}

\cabernet{} is a Java tool developed as  Cytoscape's version 3.x
application; see the Availability Section for information about
download and installation of the tool. \cabernet{} sessions are
user-defined batch computations. Parameters are specified by a
step-by-step wizard and various post-simulation functions are
accessible directly from the \cabernet{} menu in the Cytoscape active
window.  The tool implements a wide range of  simulation and analysis
functions, which can be summarily listed as follows (see
Fig.~\ref{fig:table} for a summary of functions and parameters, and
the user manual, on the plugin website, for a detailed description).
Instructions on obtaining and using the code used in the paper are
available on the page website. Tutorials include a package to
reproduce the example application discussed in the paper (see
Availability).

\subsection*{Input GRNs (generation, import and augmentation)}

\paragraph{[Random network generation]} \cabernet{} can randomly
generate and simulate ensembles of GRNs with certain structural
parameters such as: $(i)$ number of genes; $(ii)$  ingoing/outgoing
GRN's topology, e.g., {\em fixed}, \emph{Erd$\ddot{o}$s-R\'enyi's
  random} \cite{Erdos_renyi1959}, \emph{Barabasi-Alberts' scale-free}
\cite{Barabasi1999} or \emph{Watts-Strogatz's small-world}
\cite{watts1998collective}; $(iii)$ type of regulation functions
(node-specific Boolean functions). Concerning the latter, these can be
set by the user or randomly generated to accomodate:  $(i)$
\emph{bias-based random} functions, $(ii)$ \emph{canalyzing}
\cite{Kauffman2004} or $(iii)$ \emph{logical}  functions - expressed
in the canonical AND/OR notation. Network and simulation parameters
(e.g., samples size) can be defined  either via an input form or a
textual file.

\paragraph{[Import of characterized GRNs]} GRNs that are characterized
with respect to both the topology and the regulatory functions, either
fetched from public datasets via Cytoscape or loaded via input textual
file, can be processed with \cabernet{}.

\paragraph{[Augmentation of GRNs]} Any GRN loaded in the tool (see
above) can be augmented by \cabernet{}, by randomly generating a
chosen number of augmented networks, in which entities and relations
are added to the input network according to user-defined structural
and functional parameters. Parameters for augmentation are the same
that must be defined for the random generation (see above).  The
resulting ensemble of augmented networks will share the input
topological and functional core and will differ for the randomly
generated portion.

For instance, in the Results Section, the \emph{T-helper} cell
signaling network curated from \cite{klamt2006methodology} (40 genes
and 51 regulatory interactions), is augmented with 160 further nodes
and 349 edges according to a Erdos-R\'enyi random topology.
Assessment of the functional effect of this augmented network is also
discussed.

\subsection*{GRN's dynamics simulation}

Given that the space of the possible configurations of a NRBN can be dramatically large (there are $2^N$ possible configurations for a network with $N$ nodes),  \cabernet{} can simulate the dynamics of a network by either: $(i)$ sampling the initial conditions to test or $(ii)$ performing an exhaustive search (for small networks only). \cabernet{} allows to investigate key statistics of the emerging attractors such as, e.g.,  number, length, robustness and reachability. The stability of any pattern to perturbations can be assessed either  via  \emph{temporary flips} (with duration of 1 step) or via \emph{permanent} gene knock-in/knock-out.

Following the simulation,  \cabernet{} can compute and display the threshold-dependent ATN (here called the \emph{TES network}) for specific  threshold values. Different views on such a network are available in the tool and, for instance, allow to display   the genes' configurations in a  pattern, and the  variation of the number of TESs alongside thresholds, as proposed in \cite{graudenzi2014investigating}.

\subsection*{GRN-selection constrained by differentiation scheme}

One might look for a GRN giving rise to a specific differentiation
tree, as done e.g. in \cite{Graudenzi2014}. Trees can be inputed to
\cabernet{} in textual format or from the Cytoscape active window;
\cabernet{} can select those NRBNs that display a differentiation tree
structurally similar to the loaded one, where the measure of
similarity is defined by the user. This feature is implemented as  a
batch process scanning among generated NRBNs. Notice that, usually, a
single NRBN exhibits various emerging trees, according to the various
possible combinations of thresholds thus this feature is the
computationally most demanding in \cabernet{}.

Besides, the statistically \emph{representative} differentiation
tree(s) of each specific network, defined as the most frequent
emerging tree for different (sampled) threshold combinations, provided
a specific tree depth, can be computed.

\subsection*{Visualization}

Each computational task is tracked by a progress bar. Once the
simulation of the dynamics is completed, the powerful visualization
capabilities of Cytoscape can be used to analyze the topological and
dynamical properties of the networks.

In particular, with \cabernet{} it is possible to visualize: $(i)$ the
NRBNs, $(ii)$ the \emph{attractor graph network}, in which all the
states of the attractors and the transitions among them are displayed,
$(iii)$ the threshold dependent ATNs and $(iv)$ the representative
differentiation tree.  By clicking on a specific network, it gets
visualized within Cytoscape, so that it can be further analyzed.

Different network styles have been defined and can be selected: $(i)$
\cabernet{} \emph{network}, aimed at visualizing the properties of the
NRBN: the color of each node being related to the Boolean function
bias and the size of each node proportional to its degree, $(ii)$
\cabernet{} \emph{attractors}, for the visualization of the attractor
graph network, $(iii)$ \cabernet{} \emph{TES}, for the visualization
of the ATN and $(iv)$ \cabernet{} \emph{collapsed TES}, for the
collapsed visualization of the ATN: in these last two styles, the edge
size is proportional to the transition probability.

\subsection*{Robustness analysis}

Different kind of perturbations can be applied to a simulated
network. In particular, it is possible to perform a user-defined
number of $(i)$ temporary (i.e., flips) or $(ii)$ permanent (i.e.,
knock-in/knock-out) perturbations on $(i)$ a chosen number of randomly
selected nodes or $(ii)$ specific nodes. The robustness analysis can
be performed on single networks or on the whole ensemble of simulated
GRNs.

Network's stability is assessed via robustness analyses, by means of
standard measures such as \emph{avalanches} (i.e., the number of nodes
whose activation pattern is different in a perturbation experiment
with respect to the wild type scenario) and \emph{sensitivity} (i.e.,
the number of perturbation experiments in which a certain node's
pattern is affected) \cite{Serra2007,graudenzi2011robustness}. The
results of the analyses can be exported in {\tt csv} files.

\subsection*{Network analysis}

\cabernet{} offers a wide range of network-specific statistics. These
include  $(i)$ the distribution of the attractors' lengths, $(ii)$ the
basins of attraction, $(iii)$ the proportion of frozen and oscillating
nodes, plus  other classical network measures such as \emph{clustering
  coefficient}, \emph{network diameter} and \emph{average path
  length}. All the statistics can be visualized and exported; further
network measures are accessible via the network analysis tools
included in Cytoscape.

\subsection*{Outputs}

All the networks and the relative topological, functional and
dynamical properties can be exported as textual files, from both the
Wizard and the Function menu. For instance, the complete topological
and functional description of the networks can be exported so that it
can be used in simulation environments external to Cytoscape such as,
e.g., CHASTE \cite{chaste}, CompuCell3D \cite{swat12:CompuCell3D} or
the simulator described in \cite{Graudenzi2014} (see
\cite{DeMatteis2012} for a recent review on multiscale models of
multicellular systems). Also, it is possible to export the complete
description of all the attractor states, as well as information of
their basins on attraction.

\section*{Results}

Our group has  recently been focusing on the investigation of the
dynamical properties of multicellular systems via multiscale
simulations, with particular attention to the conditions that would
favor the emergence and development of tumors. To this end,
\cabernet{} was recently used to generate, simulate and visualize the
GRNs ruling the behaviour of an intestinal crypt in CHASTE's
multiscale simulation engine, allowing to identify  conditions for
cancer's emergence and  crypt's colonization \cite{cognac2015}.  In
the following, we propose a further example to show some of the
potential applications of \cabernet.

\paragraph{[Augmentation of T-helper signaling network]}
The signaling network of human T-helper cells was recently
characterized with respect to both the topology and the regulatory
functions. In \cite{mendoza2006theoretical} the dynamics of such
network was simulated with a Boolean approach and it was shown that
the attractors actually reproduce real gene activation patterns of
distinctly differentiated T-helper cell types. With \cabernet{} the
same dynamical analysis could have been easily performed.

\begin{figure}[ht!]
  \begin{center}
    \includegraphics[width=0.93\textwidth]{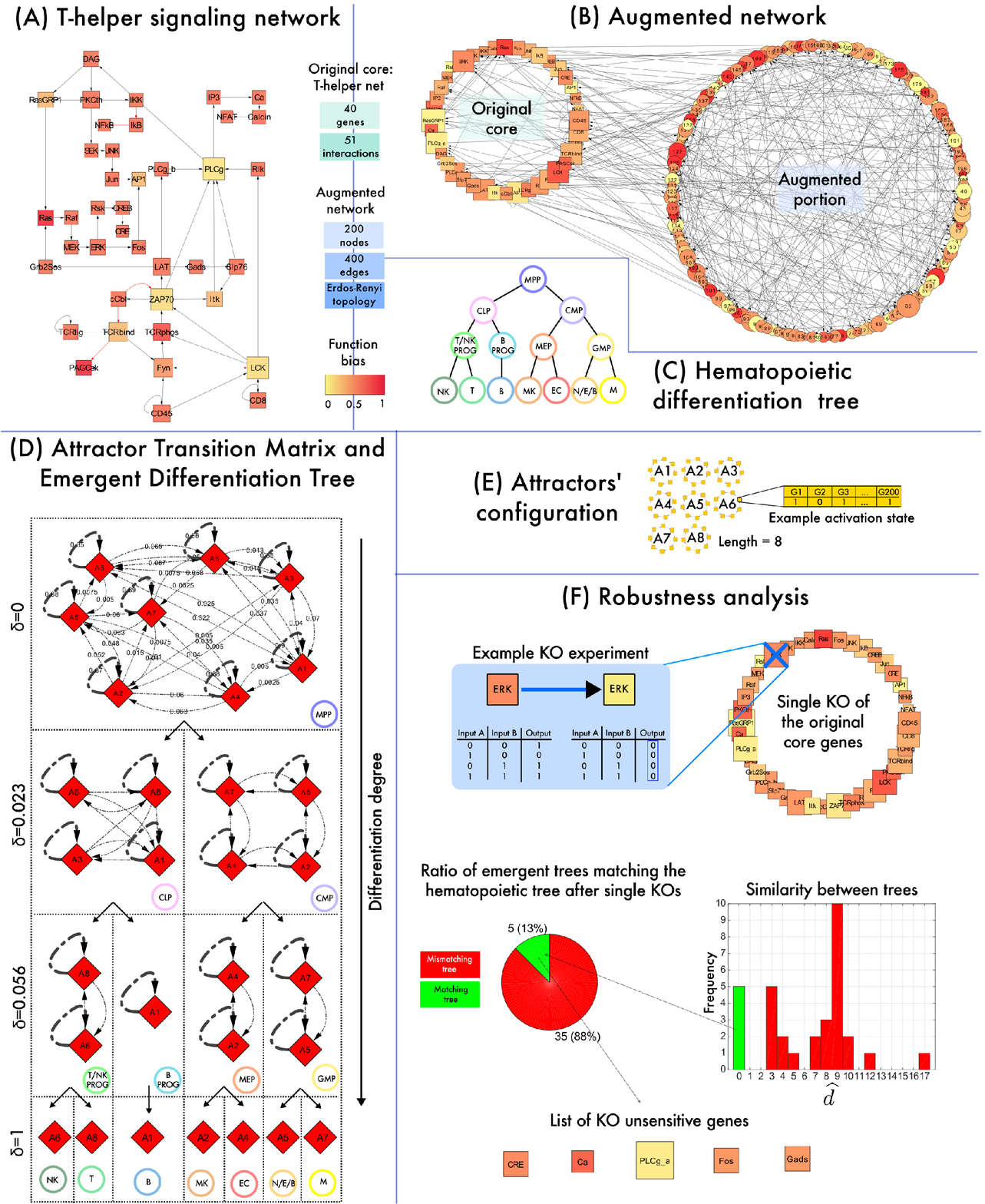}
  \end{center}
  \caption{\footnotesize{\bf Dynamical simulation and robustness
      analysis of an augmented T-helper GRN with \cabernet{}. }(A) The
    T-helper signaling network, mapped in
    \cite{klamt2006methodology}. Edges stand for regulatory
    interactions, either activating (black) or inhibiting  (red). The
    network is composed by 40 genes and 51 interactions. (B) The
    augmented NRNB that displayed a differentiation tree matching the
    hematopoietic one. To find it, 600 NRBNs were randomly generated
    by augmenting the T-helper GRN in  \cabernet{};  the augmented
    networks include 200 nodes (160 nodes added to the original core)
    and 400 edges (349 new ones, average connectivity = 2). The nodes
    are wired according to a random Erdos-Renyi topology, and random
    Boolean functions with bias = 0.5 are associated to the
    nodes. Only matching NRBN is shown,  the original core and the
    augmented portion of which are highlighted.  In  \cabernet{}'s
    visualization the size of each node is proportional to its
    connectivity degree and the color-scale to the function bias. (C)
    The differentiation tree of hematopoietic cells from
    \cite{lim2013hematopoietic} is depicted. Notice that T-helper cell
    type represents one of the leaves of the tree. For the description
    of the acronyms please refer to the main text. (D) The Attractor
    Transition Matrix of the matching NRBN is plot by \cabernet{},
    highlighting the noise-induced transitions among attractors and
    the Threshold Ergodic Sets representing cell types. The
    progressive splitting of the TESs due to increasingly larger noise
    resistance-related thresholds (i.e., $\delta = 0, 0.023, 0.056,
    1$) is shown, stressing the perfect matching between the emergent
    differentiation tree and that of hematopoietic cells, from
    multi-potent cells to fully differentiated cell types. (E)
    Configuration of the 8 attractors (determining the gene activation
    patterns). In this specific case, the length of each attractor is
    equal to 8. (F) Robustness analysis performed via
    \cabernet{}. Single node knockout experiments (i.e., silencing the
    node's Boolean function) are performed on each node of the
    original core of the augmented network and the dynamics is
    simulated again via \cabernet{}. The emergent tree is then
    compared with that of hematopoietic cells and the distribution of
    the similarity measure $\widehat{d}$ (Equation \ref{hist_dist}) is
    displayed, highlighting 5 genes that, when silenced, still lead to
    a matching emergent tree (i.e., $\widehat{d}= 0$).} 
  \label{fig:augmented}
\end{figure}

In the following, we present an experiment of GRN's augmentation
possible in \cabernet{}. To the best of our knowledge, no experiments
of this sort are possible with other RBN-based tools. The final goal
is to: $(i)$ generate a large ensemble of random networks with the
T-helper  functional core, and $(ii)$ select only those networks in
which the emergent dynamical behaviour is in accordance with the
hematopoietic differentiation tree, in which the T-helper cell type is
supposed to be one of the leaves (see Fig.~\ref{fig:augmented}).

The distinct networks differ for the augmented portion, which is
randomly generated with structural parameters (i.e., Erdos-Renyi
random topology, average connectivity $= 2$, random Boolean functions
with bias $= 0.5$) that are classically used in similar studies (see,
e.g., \cite{Serra2007,Graudenzi2011}).  Accordingly, only certain
networks will eventually display the desired emergent dynamical
behavior.  The underlying idea is that the \emph{matching} networks
might allow for the formulation of hypotheses on the missing portions
of the relevant GRN ruling the overall hematopoiesis process. Besides,
the characterization of the attractors could be matched with the real
gene activation patterns driving the functioning of the various cell
types.

In the experiment, a large number of distinct augmented networks was
generated and simulated with  \cabernet{}: only 1  on a total of 600
augmented NRBNs actually displayed the expected dynamical behavior,
i.e., the correct differentiation tree, hinting at the complexity of
the tuning process driven by the evolutionary pressure that led to the
topology of current GRNs.

In Fig.~\ref{fig:augmented} one can see the original T-helper
signaling network, originally mapped in \cite{klamt2006methodology},
and the NRBN that was selected as correct, in which the augmented
portion is highlighted. Notice that, in the augmented network, both
the topology and the functions of the original core are slightly
different from those of the T-helper GRN, as a consequence of adding
new relations linking the new and the original portions of the
net. The visualization of the network is provided via \cabernet{}, by
applying the suitable styles (see above).

In Fig.~\ref{fig:augmented} one can notice that this specific
network exhibits 8 distinct attractors (each one  characterized by a
length equal to 8 NRBN time steps). By pruning the ATN with
increasingly larger thresholds, the TES at the higher level, including
all the 8 attractors connected by noise-induced transitions and
representing multi-/toti-potent cells, progressively splits in TESs
enclosing an increasingly lower number of attractors, up to the 7 TESs
at the lower level, which correspond to single attractors, when the
threshold is equal to 1.

The resulting emergent differentiation tree perfectly matches that of
hematopoietic cells (taken from \cite{lim2013hematopoietic}, see
Fig. \ref{fig:augmented}), which is characterized by a multi-potent
progenitor (\emph{MPP}; antecedent hematopoietic stem cells, HSCs, are
not shown in the scheme), with the potential to  differentiate into
two lineages, i.e., common myeloid progenitor (\emph{CMP}) and common
lymphoid progenitor (\emph{CLP}). CMP further divide into
megakaryocyte-erythroid progenitor (\emph{MEP}) and
granulocyte/monocyte progenitor (\emph{GMP}), finally committing to
mature blood cells including erythrocytes (\emph{EC}), megakaryocyte
(\emph{MK}), monocyte (\emph{M}) and granulocytes, i.e. neutrophils,
eosinophils, basophils (\emph{N/E/B}). Conversely, CLP further
differentiate into B-cell progenitors (\emph{B PROG}) and T-cell and
natural killer cell progenitors (\emph{T/NK PROG}), with a final
commitment to mature B cells (\emph{B}), T cells (\emph{T}) and NK
cells (\emph{NK}).

Hence, it is possible to hypothesize that this specifically selected
NRBN might present topological and functional properties ensuring the
correct emergent differentiation scheme.  We remark that, in case
augmented networks were more likely to display a matching emergent
tree, one may exploit \cabernet{} to perform ensemble-level analyses
on the matching set, aimed at the formulation of hypotheses on the
generic properties of real networks.

A robustness analysis on the matching NRBN was also performed. By
simulating selective single knockouts of the genes in the original
T-helper core, we can assess the distinctive relevance in maintaining
the correct differentiation scheme. In this example, we performed 40
single {\em knockout experiments} (KO), by forcing the specific
Boolean function of each gene to inactivation (i.e., 0 output for any
regulatory input), and we tried to match the resulting differentiation
trees with that of hematopoietic cells.  Remarkably, in 35 cases
($88\%$), the KO experiment resulted in a mismatching tree, hinting at
the role of those specific genes in the interplay leading to the
emergence of the hematopoietic tree. We measured the similarity among
the hematopoietic tree, $h$,  and a tree  $T$ resulting from a KO
experiment as follows:
\begin{eqnarray*}
  \label{hist_dist}
  \widehat{d}_{h}(T)
  & = & \sum^{l^\ast}_{l=0}  \sum^{k^\ast }_{k=0}\mid n_{h}(k,l) - n_{T}(k,l)\mid
\end{eqnarray*}
where  $l^\ast$  and $k^\ast$ are the maximum depth and the maximum
number of a node's children in both $h$ and $T$. Function $n_{x}(k,l)$
returns the number of nodes at level $l$ with $k$ children in tree
$x$; thus, this quantity measures the structural {\em level-by-level
  similarity} of two trees by assessing the number of parents with $k$
children, per level. Since we focus on differentiation trees, this can
be interpreted as a measure of the ability of a certain cell type, a
progenitor,  to differentiate in a set of distinct subtypes.

In Fig.~\ref{fig:augmented} we show values of $\widehat{d}$ in our
experiments; the lower the value the closer is $T$ to $h$. Values of
$\widehat{d}$ range around 9, with a maximum of 17 and minimum 0;  in
8 cases, the value lower than 5 suggests a close similarity between
the emergent and the hematopoietic tree.  Besides, the dynamics turned
out to be completely insensitive (i.e., $\widehat{d}= 0$) to the KO of
5 specific genes, i.e., \emph{CRE}, \emph{Ca}, \emph{PLCg-a},
\emph{Fos} and \emph{Gads}, which, accordingly, might be not relevant
in the differentiation process. Clearly, further investigations are
needed to corroborate this hypothesis.

We finally remark that the generated networks could be used within any
multiscale simulation frameworks, in order to investigate, e.g., the
processes of homeostasis and clonal expansion, as proposed in
\cite{Graudenzi2014,cognac2015}.

\section*{Conclusions}
In this work we introduced \cabernet{} --  a new Cytoscape app for the
generation,  simulation and  analysis of augmented Boolean models of
gene regulatory networks --  and described some of its key
functionalities, as well as an example application to real GRN data.

\cabernet{} is the final result of a long-time effort aimed at
bridging different fields and disciplines, such as computer science,
statistics and complex systems science, for the effective study of
complex biological systems. The numerous modeling and simulation
functionalities, the various effective analysis tools and the fine
integration within the widely used Cytoscape framework, might settle
the ground for  \cabernet{} becoming a powerful instrument for
bioinformaticians and  computational biologists, especially in
providing a computational support for  experimentalists.
 
In particular, \cabernet{} can provide an essential tool to
effectively investigate key and still partially undeciphered
biological phenomena, such as, e.g., gene regulation, cell
differentiation and tumorigenesis, with particular focus on the
properties of dynamical gene activation patterns and their relation
with biological noise.

\subsection*{Availability and requirements}

{\bf Project name:} 
\cabernet{}: a Cytoscape app for the generation and the Analysis of Boolean models of gene Regulatory NETworks \\
{\bf Version:} 
1.0 \\
{\bf Plugin website:} 
\url{http://bimib.disco.unimib.it/index.php/CABERNET} \\
{\bf Operating systems:} 
platform independent \\
{\bf Software requirement:}
Cytoscape 3.x (\url{http://www.cytoscape.org/}) \\ 
{\bf Programming language:} 
Java \\
{\bf License:}  
BSD-like license (see website)

\section*{Acknowledgements}

This  project was partially supported by the ASTIL program, project
``RetroNet'', grant n. 12-4-5148000-40; U.A 053, and by NEDD Project
[ID14546A Rif SAL-7] Fondo Accordi Istituzionali 2009.

\bibliographystyle{alpha} 
\bibliography{references} 

\end{document}